%% file: cm_journal.tex
\documentclass[journal]{IEEEtran}

\usepackage{xcolor,soul,framed} 
\colorlet{shadecolor}{yellow}
\usepackage[pdftex]{graphicx}
\graphicspath{{../pdf/}{../jpeg/}}
\DeclareGraphicsExtensions{.pdf,.jpeg,.png}
\usepackage{tikz}
\usetikzlibrary{patterns,hobby,arrows,calc, positioning}
\usepackage{pgfplots}
\usepackage{adjustbox}
\pgfplotsset{compat=1.16}

\usepackage{array}
\usepackage{arydshln}
\usepackage{multirow}
\usepackage{bm}
\usepackage[cmex10]{amsmath}
\usepackage{mdwmath}
\usepackage{cite}
\usepackage{eqparbox}
\usepackage{url}
\usepackage{amsmath}
\usepackage{amssymb}
\usepackage{bm}
\usepackage{pgfplots}
\usepgfplotslibrary{statistics}
\pgfplotsset{compat=1.8}
\usepackage [autostyle, english = american]{csquotes}
\usepackage{dblfloatfix}
\MakeOuterQuote{"}
\usepgfplotslibrary{statistics}

\begin{document}
\bstctlcite{IEEEexample:BSTcontrol}

\title{Congestion Mitigation in Unbalanced Residential Networks with OPF-based Demand Management}

\author{Marta~Vanin, 
 Tom~Van~Acker, 
        Hakan~Ergun, 
	Reinhilde~D'hulst, Koen~Vanthournout
        and~Dirk~Van~Hertem.

\thanks{M. Vanin, T. Van Acker, H. Ergun and D. Van Hertem are with the Research Group ELECTA, Department of Electrical Engineering, KU Leuven, 3001 Heverlee, Belgium.

Reinhilde D'hulst and Koen Vanthournout are with VITO, Boeretang 200, 3400 Mol, Belgium.
All authors except T. Van Acker are also with EnergyVille, Thor Park 8310, 3600 Genk, Belgium.
Corresponding author: marta.vanin@kuleuven.be
}
\thanks{This work is supported by the  Flemish DSO Fluvius within the framework of the research project ADriaN – \textit{Active Distribution Networks.}}}

\maketitle

\input{sections/Abstract_keywords}

\section*{Nomenclature}
\input{sections/0_Nomenclature}

\section{Introduction}
\input{sections/1_Introduction}

\section{Mathematical Framework}
\label{maths}
\input{sections/2_Mathematical_Framework_TVA}

\section{Case study}\label{results}
\input{sections/3_Case_study}

\section{Conclusions}\label{conclusions}
\input{sections/4_Conclusions}

\section*{Acknowledgement}
\input{sections/Acknowledgments}

\bibliographystyle{IEEEtran}
\bibliography{IEEEabrv,Bibliography}

\end{document}

%% file: sections/Abstract_keywords.tex
\begin{abstract}
This paper proposes a novel congestion mitigation strategy for low voltage residential feeders in which the rising power demand due to the electrification of the transport and heating systems leads to congestion problems. The strategy is based on requiring residential customers to limit their demand for a certain amount of time in exchange for economic benefits.

The main novelty of the method consists of combining a thorough representation of the network physics with advanced constraints that ensure the comfort of residential users, in a scalable manner that suits real systems. The mitigation strategy is presented from a DSO perspective, and takes the form of contracts between users and system operator. The focus on user comfort aims to make the contracts appealing, encouraging users to voluntarily enroll in the proposed mitigation scheme.

The presented solution is implemented as a mixed-integer multi-period optimal power flow problem which relies on a linearized three-phase power flow formulation. Calculations on 100 real-life distribution feeders are performed, to analyze the congestion-relieving potential of several possible system operator-user contracts. From a planning perspective, the results can help the system operator define contractual terms that make a specific congestion mitigation scheme effective and viable. From an operational perspective, the same calculations can be used to optimally schedule power reduction on a day-ahead basis. 

\end{abstract}

\begin{IEEEkeywords}
Congestion management, demand-side management, distribution networks, flexibility, unbalanced OPF
\end{IEEEkeywords}

%% file: sections/0_Nomenclature.tex
\newcommand{\im}{{i\mkern1mu}}
\newcommand{\NaturalNumbers}{\mathbb{N}}

\newcommand{\User}{u}
\newcommand{\UserSet}{\mathcal{U}}
\newcommand{\Phase}{\phi}
\newcommand{\PhaseSet}{\Phi}
\newcommand{\Time}{t}
\newcommand{\TimeSet}{\mathcal{T}}

\newcommand{\ReductionLimit}{\eta}
\newcommand{\ActivationLimit}{\alpha}
\newcommand{\DeactivationLimit}{\delta}

\newcommand{\Status}{s}
\newcommand{\Activation}{y}
\newcommand{\Deactivation}{z}
\newcommand{\ApparentPower}{S}
\newcommand{\ApparentPowerVector}{\mathbf{\ApparentPower}}
\newcommand{\ActivePower}{P}
\newcommand{\ActivePowerVector}{\mathbf{\ActivePower}}
\newcommand{\ForecastedActivePower}{\ActivePower^{\text{fx}}}
\newcommand{\GuaranteedActivePower}{\ActivePower^{\text{gtd}}}
\newcommand{\ReactivePower}{Q}
\newcommand{\ReactivePowerVector}{\mathbf{\ReactivePower}}
\newcommand{\ForecastedReactivePower}{\ReactivePower^{\text{fx}}}
\newcommand{\GuaranteedReactivePower}{\ReactivePower^{\text{gtd}}}

This section presents the sets, parameters and variables that are used throughout the paper. Bold letters indicate matrices and vectors, lower case letters indicate either the elements of the defined sets or optimization variables. 

Let $\Phi_i$ be the set of the phases connected to a three-phase bus $i$, such that $\Phi_i = \{a,b,c\}$. Residential connections can be single- or three-phase. $\mathbf{U}_{i,t}$ is the column vector of the voltages at each phase of a bus $i$ and time $t$: $$ \mathbf{U}_{i,t} = \left[  U_{i,t,a} \; \;  U_{i,t,b} \; \; U_{i,t,c}  \right]^{\text{T}}.$$ All the vectors $\mathbf{V}$ with the same dimensions as $\mathbf{U}_{i,t}$ are synthetically written as $\mathbf{V}^{|\Phi_i|}$.

\addcontentsline{toc}{section}{Nomenclature}
\subsection*{Sets and elements}
\begin{IEEEdescription}[\IEEEusemathlabelsep\IEEEsetlabelwidth{xxxxxxxxxxxxxx}]
	\item[$t \in \mathcal{T} \subset \NaturalNumbers$] Discrete simulation time 
    \item[$ i \in \mathcal{B} $] Feeder buses
    \item[$ u \in \mathcal{U}$] Participating users, modeled as buses with power demand
	\item[$(i,j) \in \mathcal{L}$] Feeder branches
    \item[$\Phi = \{a, b, c \}$] Network phases
	\item[$ \Phi_i, \Phi_{ij} \subseteq \Phi $] Phase(s) of bus $i$ or of branch $(i,j)$
 \end{IEEEdescription} \vspace{-0.3cm} 
\subsection*{Line variables and parameters} 
\begin{IEEEdescription}[\IEEEusemathlabelsep\IEEEsetlabelwidth{xxxxxxxxxxxxxx}]
	\item[\textbf{U$_{i,t}$} $ \in \mathbb{C}^{|\Phi_i|} $] Complex voltage at bus $i$ at time $t$
	\item[$\mathbf{Z}_{ij} \in \mathbb{C}^{|\Phi_{ij}| \times |\Phi_{ij}|} $] Impedance matrix of branch $(i,j)$
	\item[$\mathbf{S}_{ij,t} \in \mathbb{C}^{|\Phi_{ij}| \times |\Phi_{ij}|} $] Apparent power flow for branch $(i,j)$ at time $t$
     \item[$\mathbf{P}^{\text{\text{fx}}}_{u,t}$  $ \in \mathbb{R}^{|\Phi_i|} $] Forecasted active power demand for user $u$ at time $t$  
        \item[$\mathbf{Q}^{\text{fx}}_{u,t}$  $ \in \mathbb{R}^{|\Phi_i|} $] Forecasted reactive power demand for user $u$ at time $t$
	\item[$\mathbf{P}_{u,t}$  $ \in \mathbb{R}^{|\Phi_i|} $] Actual active power demand for user $u$ at time $t$  
        \item[$\mathbf{Q}_{u,t}$  $ \in \mathbb{R}^{|\Phi_i|} $] Actual reactive power demand for user $u$ at time $t$
    \item[$\text{P}_u^{\text{gtd}} \in \mathbb{R}$] Guaranteed active power for user $u$
\end{IEEEdescription}

\vspace{-0.3cm} 
\subsection*{Demand reduction variables and parameters} 
\begin{IEEEdescription}[\IEEEusemathlabelsep\IEEEsetlabelwidth{xxxxxxxxx}]

	\item[$s_{u,t} \in \{0,1 \}$] Status of participating user: if user $u$ participates to the mitigation scheme at time $t$, $s_{u,t} = 1$, else, $s_{u,t} = 0$ 
	\item[$y_{u,t} \in \{0,1 \}$] Activation of a reduction action: it is 1 if a reduction action begins at $t$, 0 otherwise 
	\item[$z_{u,t} \in \{0,1 \}$] Deactivation of a reduction action: it is 1 if a reduction action stops at $t$, 0 otherwise
	\item[$\eta_u \in \mathbb{N}$] Maximum number of reduction actions per day for user $u$
	\item[$\alpha_u \in \mathbb{N}$] Maximum time length of each reduction action for user $u$
	\item[$\delta_u \in \mathbb{N}$] Minimum interval between two consecutive reduction actions for user $u$

\end{IEEEdescription}

%% file: sections/1_Introduction.tex
\subsection{Background and Motivation}

\IEEEPARstart{L}OW voltage distribution networks (LVDNs) in Europe are typically operated using a fit-and-forget approach. This is due to the relatively modest magnitude and simultaneity factor of traditional residential demand, which make the expansion of the grid capacity a simple and viable solution to overcome local issues. Furthermore, the existing infrastructure is typically sufficiently robust to operate safely and provide the required quality of service given the present demand.

However, the extensive installation of distributed generation and the electrification of the transport and heating systems increase the demand, the simultaneity factor and the congestion risk in distribution networks. In particular, the global sales of electric vehicles (EVs) and heat pumps are steadily increasing~\cite{IEA_EV, Akmal}, and their impact on the residential low voltage grid is likely to be significant~\cite{VanRoyEV}. This is because their peak demand is comparable to or exceeds the instantaneous consumption of a household, especially in the case of EVs~\cite{IEC}. Furthermore, the power consumption from these technologies coincides with the residential evening peak load~\cite{Protopapadaki}.

Therefore, in view of a future scenario with increased electrification, DSOs are facing the need for exploring strategies that allow to reduce congestion risk and ensure continuous and reliable operation~\cite{Liu}, while postponing expensive infrastructural reinforcements where possible. Thus, LVDNs are moving towards active management~\cite{Keane}.

A strategy to mitigate the congestion risk is demand-side management (DSM). For industrial users, demand reduction almost inevitably results in direct and significant economic drawbacks. In the case of residential users, there is a potential to implement such strategy at a relatively low cost or loss of comfort, as long as the guaranteed power consumption threshold is reasonable and the duration of the reduction acceptable. Furthermore, EVs and other similar, large residential loads presents can be used flexibly~\cite{Dhulst, Dar}. 
\subsection{Related Work, Contributions and Paper Organization} \label{sec:contributions}
In this paper, we propose a scalable congestion mitigation method for LVDNs that can be incorporated in a DSO's set of network management tools.

The method relies on multi-period optimal power flow (OPF)-based calculations and is based on requesting a number of residential users to keep their power consumption below a given threshold for a limited time window, when congestion risk is forecasted the day ahead. Consumers participate in exchange for economic benefits, and sign a contract that determines comfort guarantees, such as the maximum duration of power reduction and a minimum guaranteed demand threshold that is always usable.

A large number of works explore the use of optimization to address residential demand management; a comprehensive overview is given in \cite{Esther}. They can typically be divided into three categories, depending on their objective~\cite{Esther}: electricity bill minimization, user discomfort minimization and maximization of local generation use. Typically, on a residential level, the main driver for DSM methods is the minimization of the electricity cost (first category), and users are assumed to rely on a home energy management system (HEMS) that automatically controls the loads that need to be reduced or switched off~\cite{Sarker, Huang, Longethiran}. 

The method presented in this paper differentiates itself in that it belongs to the second category of problems: it aims to prevent congestion while minimizing the loss of user comfort. Furthermore, while it can be integrated in a HEMS, this is not a requirement: users who sign a contractual agreement can be timely notified about the required power limitation with day-ahead forecast results, and manage their demand accordingly. This allows reduced communication and technological overhead with respect to methods that rely on advanced HEMS, enabling the deployment of the proposed approach with today's technology. Moreover, the presented method is more inclusive than other DSM examples from the literature, where only users with storage systems \cite{Lokeshgupta}, smart appliances \cite{Anees}, EVs \cite{Lopez2015} or similar can participate. The proposed DSM method is designed to be integrated in contracts that allow direct user-DSO interaction, and could be framed as part of the amber phase in a Traffic Light-based network management approach~\cite{TrafficLights}. 

Modelling low voltage users and network physics implies a number of complications. Firstly, LVDNs present a non-negligible degree of unbalance, which requires the use of three-phase power flow equations for their realistic physical description. Secondly, the reducible loads cannot be modulated continuously; this is currently only feasible for inverter-connected devices~\cite{7, 12} or large industrial loads~\cite{1,4}. Therefore, power reduction in this work can only occur in an "ON/OFF" manner, which implies the use of binary variables in the optimization problem.

The explicit inclusion of grid voltage and thermal constraints and the binary nature of the loads results in mixed-integer nonlinear programming (MINLP) problems, and the use of the "AC" power flow formulation is NP-hard~\cite{np-hard}. Therefore, alternative strategies are typically devised to simplify the model. For example, Zhu \textit{et al.},~\cite{Zhu} and Sepulveda \textit{et al.}~\cite{Sepulveda} minimize the aggregated peak of shiftable appliances for a set of households, using integer linear programming and binary particle swarm optimization, respectively. Longethiran \textit{et al.}~\cite{Longethiran} use a heuristic-based evolutionary algorithm to perform load shifting on a greater variety of customers and load types. Zhao \textit{et al.}~\cite{ZhaoGeneticAlgorithm} use genetic algorithms to optimally schedule the demand of smart appliances minimizing the users' energy bill. Bradac \textit{et al.}~\cite{Bradac} solve a similar scheduling problem, with mixed-integer linear programming (MILP). Waseem \textit{et al.}~\cite{Waseem} use a greywolf and crow search optimization to reduce both electricity cost and peak to average ratio. All these works neglect grid constraints, and therefore there is no strict control of congestion events and voltage issues along the feeders. 

Two possible manners to tackle tractability issues while including a physical description of the system are relaxing the binary constraints~\cite{Avramidis2021} to obtain a nonlinear problem (NLP), or approximating the nonlinearities to deal with a MILP problem~\cite{Sarker}. References~\cite{Sarker, Avramidis2021} rely on balanced power flow equations, whereas the distribution system is unbalanced, and~\cite{Sarker} does not provide feasibility considerations of their MILP approximations. 

The congestion mitigation method proposed in this paper also relies on a MILP approximation of the original MINLP load scheduling problem, with the following characteristics: 
\begin{enumerate}
\item The MILP formulation used is three-phase unbalanced, which gives a better distribution system representation than balanced models (like~\cite{Sarker, Avramidis2021}), while it is still computationally tractable. Furthermore, an analysis is provided of the conditions in which the MILP solution is a feasible solution of the original MINLP problem.
\item The congestion mitigation problem is presented from the DSO perspective, exploring a number of viable possibilities that fit the contracts between system operator and consumers. For this purpose, user demand is modelled individually rather than aggregated. User participation in the contractual schemes is inclusive, i.e., not bound by ownership of EVs or similar.
\item To ensure the comfort of residential users, the proposed formulation includes advanced constraints, such as a minimum time interval between two power reduction "time windows". To the authors' knowledge, this is an original contribution to existing DSM formulations. 
\item Simulations for the proposed solution have been performed on a large set of real network and demand data.
\end{enumerate}
The combination of the four characteristics above results in a novel congestion mitigation method, which focuses on grid security and user comfort. Furthermore, the use of contractual agreements to ensure the correct LVDNs operation is under-addressed in the literature, whereas its limited technical requirements make it easy to implement in the short-term.

A simplified version of the MILP problem was presented in a previous paper \cite{Vanin2020}. The scope of \cite{Vanin2020} was to show that the used power flow formulation is suitable for mixed-integer problems like the one addressed in this work, user comfort was neglected. The basic MILP problem in \cite{Vanin2020} is here extended with user comfort constraints and is framed in a DSO-user contractual scheme which is applicable in real life.

The rest of the paper is organized as follows: Section~\ref{maths} describes the mathematical formulation of the MILP OPF problem. This is subdivided in three subsections: in~\ref{sec:user} the formulation of the user response to a load reduction request is reported, in~\ref{sec:modalities} that of the contractual constraints, and in~\ref{sec:pfe} that of the linear power flow equations.  Section~\ref{results} shows and discusses results for 100 strongly congested real-life low voltage feeders. Feeder and demand profile data have been made available by the Flemish DSO: Fluvius. The work is then summarized and concluded in Section~\ref{conclusions}.

%% file: sections/2_Mathematical_Framework_TVA.tex
\IEEEeqnarraydefcolsep{0}{\leftmargini}

This section presents a set of constraints and associated objective function to model user response to load reduction requests. Furthermore, different contract schemes are proposed, referred to as "modalities". Finally, the chosen approximation of the power flow equations is summarized. The union of these three parts constitutes the full OPF problem enabling congestion mitigation. 

It should be noted that while it is possible to extend the problem to additionally include topology reconfiguration, storage, or similar, this is not done in the present paper. The reason is that contractual agreements are easier to implement in the short-term future, given the limited or absent remote control capabilities of LVDNs, and therefore it is interesting to assess the impact of this method alone. For the same reason, power generation of users that have rooftop PV panels is included in user power profiles, but is not dispatchable.

\subsection{User Response}\label{sec:user}

The presented congestion mitigation solution is based on contractual agreements between users and DSOs, and falls into the category of incentive-based curtailable load programs~\cite{Dupont}. Participating consumers subscribed to these programs receive economic benefits for the service they provide and can be penalized or fail to receive the reward if they do not comply with their terms~\cite{Aalami}. For this reason, the present model considers that all participants always conform to the load limitation requests. Such curtailable load programs typically ask the participants to either disconnect a specific load or limit their consumption to a given threshold~\cite{Nguni}. This work is based on the latter: users can choose a guaranteed connection capacity to which they always have access to, which can be seen as an electricity tariff containing a capacity based component, and is part of the contractual agreement. If the users are not home when asked to reduce their demand, the power limit is automatically respected, as the threshold would cover at least the basic appliances' demand. In this work, it is assumed that the threshold is constant over time, as this results in simple standardized contracts. However, from a mathematical standpoint, the threshold is a scalar and can be assigned different values through the day or week without increasing the problem complexity.

A varying threshold could be adopted to mitigate possible rebound effects, together with the enforcement of a demand reduction time shift between different users. Addressing rebound effects within the proposed demand management strategy is left for future work. 

The day-ahead congestion risk is calculated using residential power consumption forecasts based on meteorological data and historic load profiles. The periods for which users need to limit their demand are communicated the day ahead, so that participants may plan their consumption accordingly~\cite{Dupont_manual}. This ensures the proposed methodology compatibility with the available state of technology: the penetration of smart meters and appliances in Flanders is currently insufficient to assume automatic or remote load control. 

To ensure that all participating users $\User \in \UserSet$ have access to their guaranteed power thresholds,~$\GuaranteedActivePower_{\User}$ and~$\GuaranteedReactivePower_{\User}$, at any discrete time period~$\Time \in \TimeSet$, a binary variable~$\Status_{\User,\Time}$ is introduced. If the user~$\User$ is not requested to reduce their power demand at a time period~$\Time$, the corresponding binary variable equals zero:~$\Status_{\User,\Time}=0$, and the power demand is assumed to be equal to the forecasted power demand: $\ForecastedActivePower_{\User,\Phase,\Time}$ and~$\ForecastedReactivePower_{\User,\Phase,\Time}$. If a reduction action is required from the user~$\User$ at a time period~$\Time$, the corresponding binary variable is set to one:~$\Status_{\User,\Time} = 1$, and the demand is assumed equal to the guaranteed power threshold:~$\GuaranteedActivePower_{\User}$ and~$\GuaranteedReactivePower_{\User}$. It is highly unlikely that a user's demand will equal the guaranteed power threshold. Nevertheless, this value is used, as it corresponds to the worst-case scenario with respect to congestions. Each user $\User$ is connected to a bus $i$, each bus hosts maximum one user and the connection cables are modeled explicitly. Thus, an injective mapping between both sets exists: $\User \in \UserSet \to i \in \mathcal{B}$, and $\ActivePower_{\User,\Phase,\Time}$/$\ReactivePower_{\User,\Phase,\Time}$ can be rewritten as $\ActivePower_{i,\Phase,\Time}$/$\ReactivePower_{i,\Phase,\Time}$. The power consumption 
of a user~$\User$ connected to bus $i$, phase~$\Phase$ at a time~$\Time$ is enforced by:
\begin{IEEEeqnarray}{ 0 l C r r }\label{eq:threshold}
    \ActivePower_{i,\Phase,\Time}   = \Status_{\User,\Time} \GuaranteedActivePower_{\User} + (1 - \Status_{\User,\Time}) \ForecastedActivePower_{\User,\Phase,\Time}, \nonumber
                                       \\ \qquad \qquad \qquad \quad
                                       \forall \User \in \UserSet, \Phase \in \PhaseSet_i, \Time \in \TimeSet: \User \to i \in \mathcal{B},         \label{eq:xxx}  \\
    \ReactivePower_{i,\Phase,\Time} = \Status_{\User,\Time} \GuaranteedReactivePower_{\User} + (1 - \Status_{\User,\Time}) \ForecastedReactivePower_{\User,\Phase,\Time}, \nonumber
                                       \\ \qquad \qquad \qquad \quad 
                                        \forall \User \in \UserSet, \Phase \in \PhaseSet_i, \Time \in \TimeSet: \User \to i \in \mathcal{B}.                                  \label{eq:yyy}
\end{IEEEeqnarray}
The resulting apparent power is summarized in vector form:
\begin{IEEEeqnarray}{ 0 l C r r }\label{eq:Sut}
    \ApparentPowerVector_{i,\Time}  &=& \ActivePowerVector_{i,\Time} + j \ReactivePowerVector_{i,\Time},
                                        & \quad \forall i \in \mathcal{B}, \Time \in \TimeSet.
\end{IEEEeqnarray}

In order to minimize user discomfort, the sum of active status variables over the feeder is minimized:
\begin{IEEEeqnarray}{ 0 l }
    \text{minimize} \sum_{\User \in \UserSet} \sum_{\Time \in \TimeSet} \Status_{\User,\Time}.            
    \label{eq:objective}
\end{IEEEeqnarray}
Table~\ref{tab:user_scenarios} summarizes all four possible scenarios for a status variable~$\Status_{\User,\Time}$ given feeder congestion and the user's forecasted consumption. It should be noted that given \eqref{eq:objective}, a status variable~$\Status_{\User,\Time}$ is only subject to change whenever the feeder is congested and the forecasted consumption of a user~$\User$ exceeds the guaranteed power at a time period~$\Time$.

\renewcommand{\arraystretch}{1.5}
\begin{table}[b!]
    \centering
    \caption{Values of~$\Status_{\User,\Time}$ for a given forecast and feeder congestion.}
    \label{tab:user_scenarios}
    \begin{tabular}{c:c|cc}
    \multicolumn{2}{c |}{}                                  & \multicolumn{2}{c}{$\ForecastedActivePower_{\User,\Phase,\Time} >         \GuaranteedActivePower_{\User}$}  \\ \cdashline{3-4}
    \multicolumn{2}{c |}{}                                  & No                                    & Yes                                   \\ \hline
   \multirow{2}{*}{\rotatebox[origin=c]{90}{Cong.}} &   No  & $\Status_{\User,\Time} = 0$   & $\Status_{\User,\Time} = 0$   \\
                                                    &   Yes & $\Status_{\User,\Time} = 0$   & $\Status_{\User,\Time} = \{0,1\}$                             \\
    \end{tabular}   
\end{table}
\renewcommand{\arraystretch}{1.0}

\subsection{Contract Modalities}\label{sec:modalities}

Additional comfort guarantees for a user~$\User$ may be ensured through contractual terms, including:
\begin{enumerate}
    \item [(a)] maximum number of reduction actions per day:~$\ReductionLimit_{\User}$,
    \item [(b)] maximum reduction duration:~$\ActivationLimit_{\User}$, and
    \item [(c)] minimum interval between reduction actions:~$\DeactivationLimit_{\User}$.
\end{enumerate}
These comfort guarantees are mathematically realized through two additional binary variables. The activation and deactivation of a reduction action are described by~$\Activation_{\User,\Time}$ and~$\Deactivation_{\User,\Time}$, respectively. The relative behavior between variables~$\Status_{\User,\Time}$, $\Status_{\User,\Time-1}$, $\Activation_{\User,\Time}$ and~$\Deactivation_{\User,\Time}$ is governed by:
\begin{IEEEeqnarray}{ l C r r }
    \Status_{\User,\Time} - \Status_{\User,\Time-1} + \Activation_{\User,\Time} - \Deactivation_{\User,\Time} &=&0,
            &\quad \forall \User \in \UserSet, \Time \in \TimeSet \setminus\{0\}.
    \label{eq:relative_behavior_modalities}
\end{IEEEeqnarray}
Three additional constraints are introduced to ensure comfort guarantees (a)-(c). The presented formulation is inspired by~\cite{tight}, to which the maximum reduction duration constraint~\eqref{eq:maximum_redunction_duration} is added. The proposed formulation is the tightest possible way of formulating these comfort guarantees. Firstly, comfort guarantee~(a) is ensured by limiting the number of reduction activations for a specific user~$\User$ to~$\ReductionLimit_{\User}$:
\begin{IEEEeqnarray}{ 0 l C r r }
    \sum_{\Time \in \TimeSet} \Activation_{\User,\Time} &\leq& \ReductionLimit_{\User},
            &\quad \forall \User \in \UserSet.
\end{IEEEeqnarray}
Secondly, comfort guarantee~(b) is enforced by ensuring that the sum of the status variables~$\Status_{\User,\Time}$ over any time frame of length~$\ActivationLimit_{\User}+1$ is lower or equal to~$\ActivationLimit_{\User}$:
\begin{IEEEeqnarray}{ 0 l C r r }
    \sum_{\Time^{*} \in \TimeSet^{\text{*}}(\Time,\ActivationLimit_{\User})} \Status_{\User,\Time^{*}} &\leq& \ActivationLimit_{\User},
            &\quad \forall \User \in \UserSet, \Time \in \TimeSet,
    \label{eq:maximum_redunction_duration}
\end{IEEEeqnarray}
where~$\TimeSet^{\text{*}}(\Time,x) = \{\min(0,\Time-(x+1)),...,\Time\}$.

Thirdly, comfort guarantee~(c) is enforced by ensuring that the sum of deactivation variables~$\Deactivation_{\User,\Time}$ over any time frame of length~$\DeactivationLimit_{\User}+1$ cannot exceed the complement of the status variable~$1-\Status_{\User,\Time}$ at the end of that time frame:
\begin{IEEEeqnarray}{ 0 l C r r }
    \sum_{\Time' \in \TimeSet^{\text{*}}(\Time,\DeactivationLimit_{\User})} \Deactivation_{\User,\Time'} &\leq& 1 - \Status_{\User,\Time},
            &\quad \forall \User \in \UserSet, \Time \in \TimeSet.
            \label{eq:deactivation_limit}
\end{IEEEeqnarray}

The combination of constraints \eqref{eq:relative_behavior_modalities}-\eqref{eq:deactivation_limit} ensures that activation and deactivation of a reduction action do not occur simultaneously.

\subsection{Power Flow Equations}\label{sec:pfe}

The inclusion of a large number of binary variables, in addition to the need to solve the OPF problem in a reasonably short time, calls for the use of a linear approximation of the power flow equations. The use of the exact ("AC") power flow formulation would make the problem nonconvex and NP-hard~\cite{np-hard}, and mixed-integer applications proved untractable with present MINLP solvers~\cite{Jay}.

The approximation of the power flow equations proposed by Gan and Low~\cite{GanLow} is chosen, which is a generalization of the simplified \textit{DistFlow} equations. It allows to model radial three-phase grids taking into account resistive losses, reactive power and voltage sag, which enable a realistic representation of distribution networks. A comparison of power flow formulations is performed in \cite{Vanin2020}, which shows that the chosen linearization provides very accurate results for LVDNs. 

For convenience, the linear formulation implemented in the OPF tool is summarized here. For its complete derivation, the reader is referred to \cite{GanLow}. 

Firstly, the complex voltage variable $\mathbf{U}_{i,t}$ is replaced by:
\begin{equation}\label{eq:Ulift}
\mathbf{u}_{i,t}= \mathbf{U}_{i,t} \mathbf{U}^{\text{H}}_{i,t}, \; \; \forall i \in \mathcal{B}, t \in \mathcal{T},
\end{equation}

where the $(\cdot)^{\text{H}}$ indicates the hermitian transpose. Let's indicate $ \mathcal{B}^+$ as the set of all the buses connected to bus $j$, for which the direction of the power flow is $ i \rightarrow j \;  \forall i \in \mathcal{B}^+ $ and $\mathcal{B}^-$ the set of all the buses connected to bus $j$, for which the direction of the power flow is $ j \rightarrow k \; \forall k \in \mathcal{B}^- $.
The assumption that power losses along the lines are limited allows to write the power balance equation as:

\begin{equation}  \label{eq:S}
\sum_{i \in \mathcal{B}^+} \text{diag}(\mathbf{S}_{ij,t}) + \mathbf{S}_{j,t} = \sum_{k \in \mathcal{B}^-} \text{diag}(\mathbf{S}_{jk, t}), 
\end{equation}
where $\mathbf{S}_{j,t}$ is the power injection at bus $j$ and time $t$ \eqref{eq:Sut}.

Assuming that voltages are nearly balanced, the off-diagonal entries of $\mathbf{S}_{ij,t}$ can be approximated as follows. Let's define $\mathbf{\Lambda}_{ij,t} = \text{diag}(\mathbf{S}_{ij,t})$ and let $\text{diagm}(\bm{\Lambda}_{ij,t})$ denote a 3$\times$3 diagonal matrix with diagonal $\bm{\Lambda}_{ij,t}$, then:
\begin{equation}\label{eq:Sij_new}
 \mathbf{  S }_{ij,t} =  \bm{\gamma} \cdot \text{diagm}(\bm{\Lambda}_{ij,t}),
\end{equation}
where $\bm{ \gamma }$ is defined as:
\begin{equation}
\bm{\gamma} = \begin{bmatrix} 
1 & \alpha^2 & \alpha \\
\alpha & 1  & \alpha^2 \\
\alpha^2 & \alpha & 1
\end{bmatrix},
\end{equation}
with $\alpha = e^{-i2\pi / 3}$. Thus, \eqref{eq:S} can be re-written as:
\begin{equation}
\sum_{i \in \mathcal{B}^+} \mathbf{\Lambda}_{ij,t} + \mathbf{S}_{j,t} = \sum_{k \in \mathcal{B}^-} \mathbf{\Lambda}_{jk,t}. 
\end{equation}
The Ohm's law, which relates power flow to voltage differences becomes:
\begin{equation} \label{eq:v}
\mathbf{u}_{j,t} = \mathbf{u}_{i,t} -\mathbf{S}_{ij,t}\mathbf{Z}_{ij}^{\text{H}} - \mathbf{Z}_{ij} \mathbf{S}_{ij,t}^{\text{H}},
\end{equation}
with $\mathbf{S}_{ij,t}$ from \eqref{eq:Sij_new}.

Finally,  $\underbar{$U$}$, $\overline{U}$ and $\overline{S}_{ij} \in \mathbb{R}$ represent respectively the voltage magnitude and thermal limits of the feeder, such that:
\begin{equation}\label{eq:Ulim}
\underbar{$U$}^2 \leq |u_{i,t,\phi}| \leq \overline{U}^2  \hspace{0.5cm} \forall i \in \mathcal{B}, t \in \mathcal{T}, \phi \in \Phi_i
\end{equation}
\begin{equation}\label{eq:last_opf_equation}
|S_{ij,t,\phi}| \leq \overline{S}_{ij}  \hspace{0.5cm} \forall (i, j) \in \mathcal{L}, t \in \mathcal{T}, \phi \in \Phi_{ij}.
\end{equation}

Equations \eqref{eq:threshold} to \eqref{eq:last_opf_equation}, with \eqref{eq:objective} as objective, form the full multi-period OPF problem.

%% file: sections/3_Case_study.tex
The DSO can use the proposed OPF method in both the planning and operation phase of their networks. In the planning phase, calculations are performed offline, to assess which socially acceptable contractual modalities are effective in relieving congestion. Once a satisfactory modality is found, these can be used at the operational level, to calculate the next day's power reduction schedule.

The grid data and samples of demand profiles used in this section are made available by the Flemish DSO, Fluvius. A model of a future large EV fleet is added to the demand profiles, to induce congestion and power quality issues in the otherwise robust present grids. The EV model is based on the current traditional mobility behaviour in Flanders~\cite{OVG3}. Details on the case study are presented in Section~\ref{sec:csdescription}. Although EVs have been chosen as the "congestion-generating" technology in this work, the use of any other device or combination of devices would not have an impact on the problem formulation itself. What might change is the effectiveness of the different contractual modalities: the absolute results are strongly dependent on the scenario and the calculations should be repeated if this differs.

Section~\ref{sec:MINLP} shows that using the full AC power flow equations to solve the proposed OPF problem would result in a computationally intractable MINLP, justifying the MILP approximation. Section~\ref{sec:ACfeas} shows that the MILP problem returns more AC feasible solutions, as long as some requirements are met. 

Finally, Section~\ref{sec:modality_comparison} presents an example of the application of the proposed method in the planning phase.
The results on computational time show that the MILP problem can be solved reasonably fast and can be adopted in operations.

The problem implementation is based on PowerModelsDistribution.jl \cite{TPPM}, an open source software package in Julia/JuMP \cite{JuMP} that features different power flow formulations \cite{powermodels}. All calculations are performed on a 64-bit machine with Intel(R) Xeon(R) CPU E5-4610 v4 @1.80GHz and 32 GB RAM.

\subsection{Description of the Case Study}\label{sec:csdescription}
In this proof of concept, five different power reduction modalities have been examined:

\begin{enumerate}
\item Simple: any participating user can be asked to reduce their consumption for an unlimited number of times per day, of unlimited duration.
\item Single: any participating user can receive maximum one request per day, of maximum 6 hours.
\item Double: any participating user can receive maximum two requests per day, for a reduction duration of maximum 3 hours each.
\item Double w. $\delta$: as double, with the addition of an interval of at least 3 hours between the requests.
\item Triple w. $\delta$: maximum three 2 hours reductions per day, with an interval of at least 2 hours between any of them.
\end{enumerate}
\begin{table}[b!]
\centering
\caption{Parameters for each modality}
\label{tab:parval}
\begin{tabular}{l c c c} 
\hline
Name & $\eta$ & $\alpha$ & $\delta$\\
\hline
Simple & $\infty$ & $\infty$  & 0 \\
Single & 1 &6 hrs & 0  \\
Double & 2 &3 hrs  & 0  \\
Double w. $\delta$ &2 & 3 hrs & 3 hrs  \\
Triple w. $\delta$ & 3& 2 hrs & 2 hrs  \\
\hline 
\end{tabular}
\centering
\end{table}
The parameter values are summarized in Table \ref{tab:parval}. It is assumed that the DSO is interested in exploring a series of standardized contracts, in which the parameters are the same for all participants. Thus, $\alpha_u, \eta_u, \delta_u$ are generically replaced by $\alpha, \eta, \delta$.

The simple modality presents no guarantees for user comfort, but serves as a benchmark to compare the other four modalities: this type of modality returns the largest feasible solution space for congested scenarios, while the others might have no feasible solution due to the additional constraints. 

\begin{table}[b!]
\centering
\caption{Specifications of the case study}
\label{tab:scenario}
\begin{tabular}{l c} 
\hline \\
Number of analyzed feeders & 100\\
Customer number per feeder & 11 - 101 \\
Total number of customers & 3640 \\
EV charging power & 3.3 kVA, single-phase \\
Households with an EV & 30 \% \\
Potential participating users & All, regardless of EV ownership \\
Simulated day & 17/01/2016 \\
Time step resolution & 15 minutes \\
\hline 
\end{tabular}
\centering
\end{table}

Table~\ref{tab:scenario} reports the main features of the performed simulations. The simulated time period is 24 hours, as the intention is to make a schedule for the following day. The chosen day is the coldest day of 2016 in Belgium, which typically corresponds to the day with the highest residential power consumption (worst-case scenario). The resolution of the data from the power profiles database is 15 minutes. Thus, a full day corresponds to 96 time steps.

In the present analysis, it is assumed that the demand forecast is exact. This is because this paper addresses the planning stage of the contracts, where the effectiveness of the MILP formulation and the congestion mitigation potential of the different modalities is being examined. Hence, the decision to perform an assessment on a worst-case scenario like that of Table~\ref{tab:scenario}. Probabilistic considerations are out of scope here, but should be taken into account when addressing the operational stage of the proposed method.

Finally, it is assumed that all users of a feeder can enrol in a power reduction scheme. In this way, it is possible to deduce the average amount of required participants to make each modality possible. 

\subsection{Tractability of MINLP problem}\label{sec:MINLP}

To examine the difference in computational effort between the MINLP and the MILP problems, the "simple" modality from Table~\ref{tab:parval} is tested on all 100 feeders. The MINLP program is obtained by implementing the mixed-integer power reduction scheme to the full AC polar formulation of the power flow equations available in~\cite{TPPM}. Due to the largely increased calculation times for the MINLP, only 24 time steps are examined (6 hours), instead of a full day.

To solve the MINLP problem, Juniper 0.5.3~\cite{juniper} is used, with Ipopt 3.12.10~\cite{ipopt} with the HSL MA27 subroutine~\cite{HSL} and Gurobi 9.0.1 as underlying nonlinear and MIP solvers, respectively. This allows to use the feasibility pump feature of Juniper. Gurobi is used for the MILP problem throughout the paper. A solver time limit of 1 hour per feeder is set.

It should be noted that the existence of a solution is always guaranteed with the "simple" modality, as there is no limitation on power curtailment.

In 68~\% of the feeders, even after reducing the problem to 6 hours, the solver fails to find a solution for the MINLP problem before the time limit is reached. In the MILP case, a solution for the same feeders and problem is always found in less than 2.8~s, and on average in 0.75~s.

In the remaining 32 feeders, it takes on average 440~s to find a solution for the MINLP problem, while the MILP case is typically 3 orders of magnitude faster.

Let $O^{MINLP}, O^{MILP}$ be the objective of the MINLP and MILP problems and let $\beta$ be the maximum number of binary variables. The relative objective error $\epsilon$ can be defined as:
\begin{equation}\label{eq:rel_err}
    \epsilon = \frac{ | O^{MILP} - O^{MINLP} |}{\beta} 
\end{equation}
In 17 of the 32 solved feeders, $\epsilon$ = 0. In 12 of them,~$ 0 < \epsilon \leq 0.5\%$. In the 3 remaining cases, $ 2 \% < \epsilon \leq 4.2 \%$.

Even though the number of time steps is reduced to one fourth of that of the original problem, finding solutions for the MINLP problem takes too long for operational purposes and is also impractical for planning. On the other hand, the MILP problem is tractable and its objective values are similar to those of the MINLP problem. 



\subsection{AC feasibility of MILP solution}\label{sec:ACfeas}

Section~\ref{sec:MINLP} shows that the MINLP is extremely slow to solve, which makes the MILP approach appealing. The aim of this section is to analyze whether the MILP solutions also effectively result in congestion-free scenarios. 

A congestion event occurs whenever a bus or a branch presents an undervoltage or overcurrent, exceeding the limits~\eqref{eq:Ulim}-\eqref{eq:last_opf_equation}.

To assess AC feasibility, the reduced power scheduled with the MILP method is used as input of an AC power flow problem, similar to~\eqref{eq:Ulift}-\eqref{eq:last_opf_equation}.

The analysis shows that only 9\% of the MILP solutions are immediately AC feasible. However, if the bounds on~\eqref{eq:Ulim} and~\eqref{eq:last_opf_equation} are tighter in the MILP problem than in the original AC power flow, it can be made sure that all congestion events are avoided. Bound tightening has recently been proved an effective method to bridge the gap between solutions from approximated power flow models and exact models~\cite{Bent, Jay}.

\begin{figure}[!h]
   \includegraphics[width=3.5in]{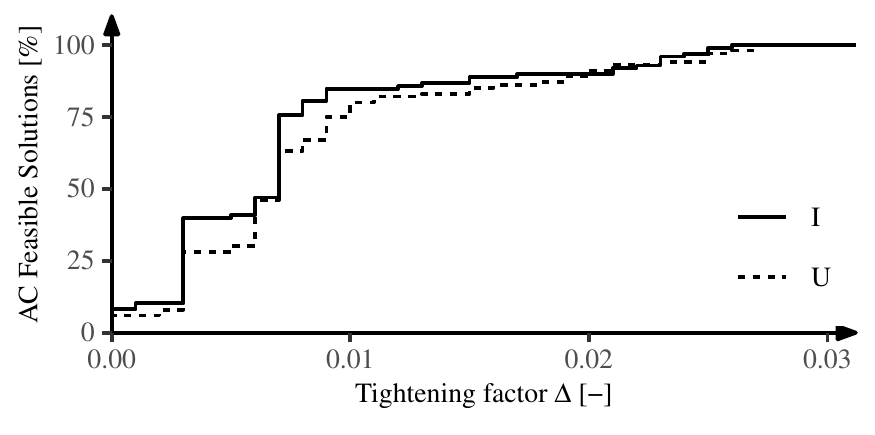}
   \centering
   \caption{Percentage of feeders solved with the "simple" modality of MILP that are completely congestion-free according to the AC PF check, in function of the restriction on voltage and power limits.}
   \label{fig:cml_simple}
\end{figure}

Fig.~\ref{fig:cml_simple} illustrates the increase of AC feasible feeders when iteratively tightening the bounds of the "simple" modality MILP. The x-axis reports the $\Delta$ parameter, which indicates the amount of tightening, as per the following definition. In feeders with overcurrent problems:
\begin{equation}
   \overline{S}_{ij, MILP}= \overline{S}_{ij,AC OPF} \cdot (1-\Delta) \; \; \; \forall (ij) \in \mathcal{L} 
\end{equation}

In feeders with undervoltage problems: 
\begin{equation}
   \underline{U}_i^{MILP}= \underline{U}_i^{ACOPF}-\Delta \; \; \; \forall i \in \mathcal{B}. 
\end{equation}

In feeders with both undervoltage and overcurrent problems, the maximum of the two $\Delta$ above is reported. Fig.~\ref{fig:cml_simple} shows that in more than 70\% of the cases, a reduction of less than 0.01 p.u. is enough to ensure AC feasibility. Increasing $\Delta$ to 0.027 p.u. ensures feasibility in 100\% of the cases.

This type of feasibility analysis shows that the MILP approximation is effective at capturing the physics of the original problem. It is then up to the system operator to decide whether to pursue a feasibility rate of 100\% or settle on lower $\Delta$ values that prove to still achieve the desired results in practice. 

The feeders used in the present analysis are pushed into strong congestions due to the added EVs. In general, the more congested the feeder, the larger the limit restriction required to ensure AC feasibility. It should be noted that while the "simple" modality allows to solve all scenarios (there are no limitations on the number of power reduction actions), highly congested feeders would not have a MILP solution for the other modalities, due to their extra constraints, as shown in the next section. 

\subsection{Modality comparison}\label{sec:modality_comparison}

This section analyzes simulation results for the modalities described in section~\ref{sec:modalities}, providing an example on how to choose the most ideal contractual agreements in the planning phase. The simulations provide the following information:
\begin{itemize}
\item how many feeders can be relieved from congestion using contractual agreements as the only mitigation solution,
\item which contractual parameters allow the least total demand reduction per participant,
\item how many users need to participate in a flexibility scheme to make it viable, and
\item which modalities imply more computational effort.
\end{itemize}

Table~\ref{tab:MILPfeas} reports the percentage of feeders for which congestion issues are completely solved with the proposed DSM approach as the only mitigation method. It can be observed that, in general, the more elaborate the user comfort constraints are, the lower the congestion relief capabilities are.

Regardless of the chosen modality, at least 34\% of the congested feeders can be relieved with the proposed method only, without the need for reinforcement. It should be noted that all the numerical results presented in this section are strongly influenced by the feeder characteristics and by the demand profiles. In particular,The high number of EVs, combined with an exceptionally cold day, result in worst case conditions that make the feeders particularly congested.

\begin{table}[t!]
\centering
\caption{Percentage of feeders in which the MILP problem has a solution}
\label{tab:MILPfeas}
\begin{tabular}{c c c c c} 
\hline
Simple & Single & Double & Double w. $\delta$ & Triple w. $\delta$\\
\hline
100\% & 42\% & 44\%  & 36\% & 34\% \\
\hline 
\end{tabular}
\centering
\end{table}

\begin{figure}[b!]
\begin{center}
\includegraphics[width=0.5\textwidth, angle=0]{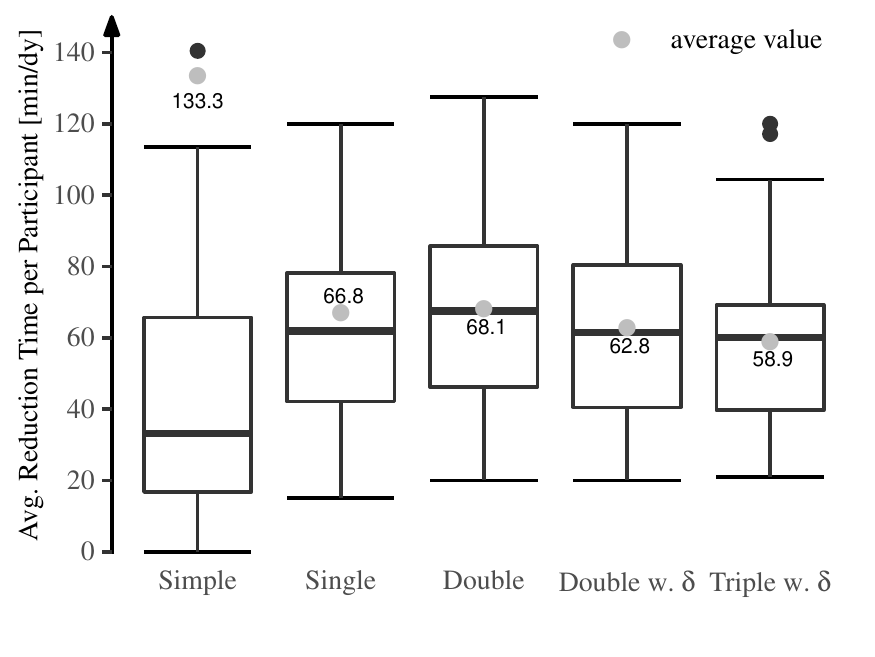}
\caption{Distribution of the average power reduction per participant over the feeder.}
\label{fig:objective}
\end{center}
\end{figure}

Fig.~\ref{fig:objective} shows the distribution of the average power reduction per participant in each feeder. The results are quite similar, except for the "simple" modality. This is because extremely congested feeders are only relieved with large reduction, which exceeds the constraints with the other modalities. For this modality, 5 outliers are not reported in Fig.~\ref{fig:objective}, for ease of representation. Their values are approximately 377, 700, 1023, 1231, 1288 min/participant.

\begin{figure}[b!]
\begin{center}
\includegraphics[width=0.5\textwidth, angle=0]{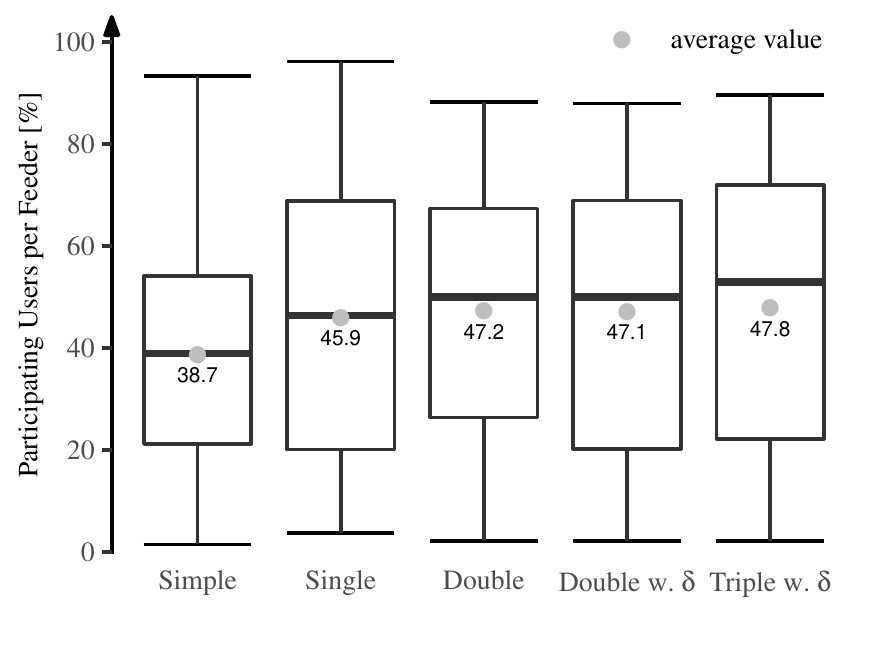}
\caption{Variations in the percentage of required participating users.}
\label{fig:users}
\end{center}
\end{figure}

Fig.~\ref{fig:users} shows the percentage of users in the feeder that needs to participate to make the modalities effective. It should be noted that these value are "ideal", as the optimizer chooses the most convenient users. Thus, in real-life, the actual number of required participants will likely be higher, depending on how the contractual agreements are realized in real life. A possibly interesting alternative for the system operator could be to repeat the present exercise minimizing the number of participants as an objective, should this prove to be a bigger bottleneck to the implementation of the contractual schemes than the required power reduction.

\begin{figure}[h!]
\begin{center}
\includegraphics[width=0.5\textwidth, angle=0]{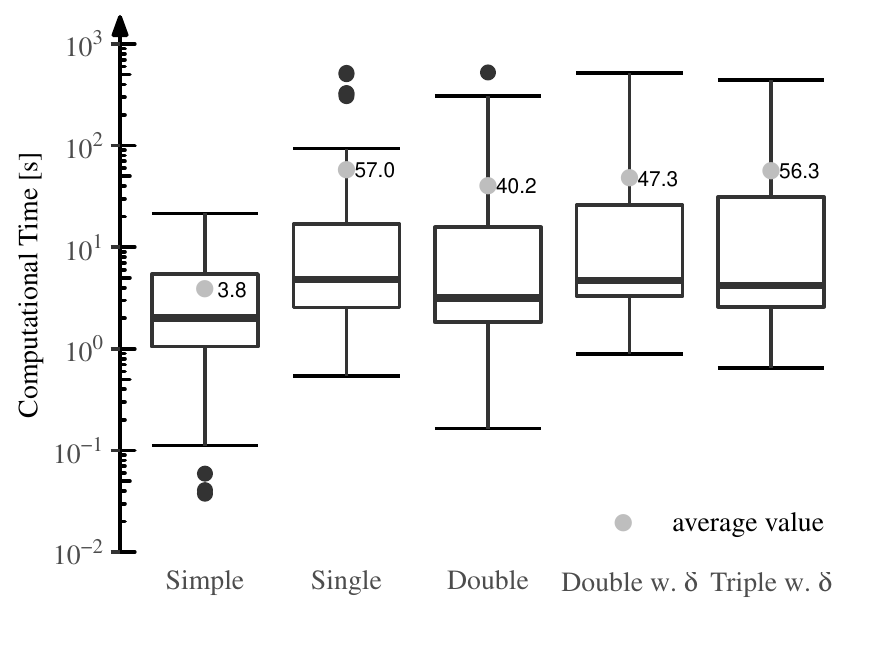}
\caption{Variation of the computation time for the different modalities.}
\label{fig:solvertime}
\end{center}
\end{figure}

Finally, Fig.~\ref{fig:solvertime} shows the time required by the solver to find a solution. Regardless of the modality, the solver times seem acceptable to schedule demand reduction on a day-ahead basis. 

The results in Table \ref{tab:MILPfeas} and Fig. \ref{fig:objective}-\ref{fig:solvertime} help the DSO decide on the most suitable contractual agreements. For example, the "Triple w. $\delta$" modality requires the highest average number of participants and presents the lowest percentage of feeders with solution, and thus is not an effective scheme.

It is up to the DSO, together with the regulator and policy makers, to assess which of the examined features are most critical and decide on the most appropriate modality. The numerical results provided in this paper are only intended as an example, and they are subjected to large variations depending on the feeder topology and the power demand pattern. The proposed method seems promising even for the strongly congested scenarios used for this paper, as at least one third of the feeders are relieved from congestion on the worst-case day of the year, regardless of the modality, without reinforcing the infrastructure.

%% file: sections/4_Conclusions.tex
Low voltage networks are accomodating increased amounts of electrically powered heating and transport equipment. In the future, these may require a strong increase in network reinforcement to avoid the congestion.

This work presents a strategy to prevent congestion in low voltage residential feeders by means of contracts between residential users and DSO, as an alternative to reinforcement.
Users who sign these contracts agree to keep their demand below a certain threshold during a given amount of time, when congestion risk is forecasted. Constraints on the power limitation modalities are introduced to maintain user comfort. These constraints are represented by a set of parameters that define standard contractual agreements. Users can stipulate such kind of contracts, trading the flexibility they offer for economic benefits. 

The proposed congestion mitigation strategy can be modelled as a mixed-integer multi-period OPF problem, and calculations can be performed in a planning phase to assess which sets of contractual parameters are the most suitable for a given low voltage system. Once the ideal set of parameters is found, the same calculations can be repeated to schedule the reduction actions on a day-ahead basis, in the operation phase.

Results for 100 real Flemish low voltage feeders show that the OPF problem is not tractable using non-convex power flow equations, whereas the linearization adopted in this work does converge in acceptable time and returns solutions that are AC-feasible, if voltage and current bounds are slightly adjusted. Finally, a proof of concept of the proposed method is presented, in which a set of contracts is tested on strongly (artificially) congested feeders. 
The results show the potential of using such contractual schemes to relieve congestions, and can serve as means of investment deferral in LVDNs. In the specific case of the examined contracts, the number of required participants does not vary significantly in the different modalities. Finally, more complex modalities seem less effective in relieving congestion issues, but when they do they require lower demand reduction from the participants. 

%% file: sections/Acknowledgments.tex
The authors would like to thank Joris Lemmens and Vincent Vancaeyzeele from Fluvius for the valuable discussions and contributions. 